# A DFT + U Study of Absorption Spectra and Localized Surface Plasmon Resonances of few electrons in Doped ZnO Quantum Dots

D Dada and M D Mochena*
Florida A&M University, Tallahassee, Florida 32307

Abstract

Quantum plasmonics of few electrons that was reported a few years ago in photodoping experiments has received little or no attention in doped nanomaterials both theoretically and experimentally. There are no studies of quantum plasmonics of large quantum dots of the size of Bohr excitonic radius or less that take electronic and geometric structure of the quantum dots. In this work, we studied extensively the absorption spectra of a protype quantum dot of sizeable ZnO of approximately 3.0 nm in diameter doped with Ga, and Al in diluted limit with density functional theory (DFT) plus Hubbard corrections (U). The localized surface plasmon resonances (LSPRs) were then determined from the real part of the dielectric function by correlating the negative portion of it to resonant spectral lines. Our results show the sensitivity of the spectral lines to distribution of the dopants, the electronic structure of the dopants, the polarization of the electric field, and size of the quantum dots. Even though our findings are based on stoichiometrically simple ZnO, it shows that the DFT + U method with numerical atomic basis can be used at reasonable computational cost to study quantum plasmonics of few electrons taking into account electronic and geometric structures, which is missing currently. The method can be extended to study magneto-optics of diluted magnetic semiconductor QDs.

## I. Introduction

Quantum plasmonics of nanostructures emerged as a topic of great research interest more than a decade ago because of the fundamental physics involved and its potential in realization of quantum-controlled devices.[1,2] The interaction of light at nanoscale with electron charge density results in surface plasmons that propagate as charge density waves at metal – dielectric interface or localized on nanostructures resulting in confinement of light to scales far below conventional optics.[3,4] Several potential applications of the extreme concentration of light have been suggested including nanophotonic lasers and amplifiers,[2,5,6] optical metamaterials,[7] biochemical sensing[8] and antennas transmitting and receiving light signals at the nanoscale.[2,9] Single – photon sources, transistors, and ultra-compact circuitry have also been proposed.[1,2]

Despite advances in quantum plasmonics, there are some outstanding fundamental issues that remain. For instance, surface plasmons are quanta of oscillations of large density of electrons according to conventional wisdom. It is then no wonder that metals and highly doped semiconductors are the ones that have been researched extensively. Recent findings from photodoping experiments, however, of resonant collective oscillations, known as LSPRs, involving as few as three electrons by one experimental group[10] and four electrons by another one[11] have raised fundamental questions about the nature of localized surface plasmons.[12]

Apart from photodoped semiconductor nanocrystals, optical transitions in artificial atoms consisting of one to ten electrons occupying the conduction levels in zinc oxide nanocrystals were studied in which the electron number was controlled through electrochemical potential.[13, 14] The major challenge in such studies was noted to be the fabrication of devices in which the electron number is controlled. To our knowledge, although several experimental studies of highly doped nanocrystals have been reported, we have not come across any studies involving few electrons in lightly doped QDs of sizes equal to or less than the excitonic Bohr radius. On theoretical front, there has been an attempt to qualitatively extend to limit of few electrons the calculations of time dependent density functional theory (TDDFT) for highly doped nanocrystals in which the discreteness of the nanocrystal was replaced by a jellium background positive charge distribution.[15] More recently, a model Hamiltonian has been applied to study the frontier between classical and quantum plasmonics of highly doped semiconductor layers with an arbitrary one dimensional potential.[16] It is then fair to assert that in quantum plasmonics, in general, both experimental or theoretical studies that probe detailed geometric and electronic structure of the LSPR – active QDs to fine tune their properties are lacking.[17] So far, the focus has been in understanding the evolution of LSPRs treating the conduction electrons as electron gas confined to one – dimensional or three – dimensional box or using TDDFT with electrons embedded in jellium positive background charges. The discreteness of the host lattice and dopants as well as the surface morphology of the nanocrystals have not been considered. In dilute limit, however, these factors cannot be neglected. In this work, we investigate if LSPRs of few electrons could result in donor – doped QDs taking into account the electronic and geometric structures of the doped QDs.

Metal oxide nanocrystals constitute a major class of plasmonic semiconductor nanomaterials. It has been pointed out understanding the evolution of LSPRs in doped metal oxide nanocrystals is one of the remaining challenges in quantum plasmonics.[17,18] The physics and chemistry of plasmonics in metal oxide nanocrystals is complex and requires a simple system, at least initially, to understand the interplay among the various factors affecting the emergence of LSPRs. It involves oxygen deficiencies, surface morphology, ionized impurities in the case of doped nanocrystals (NCs) as well as electron – electron correlation. We will not take intrinsic defects into account in this work; we will investigate a pristine system first to get a basic understanding or chemical trends. Among the oxides, ZnO, with its large band gap of 3.37 eV and relatively easier stoichiometry, for instance compared to $In_2O_3$, is a protype candidate to study LSPRs in doped QDs. Currently, there is a great deal of interest to tune LSPRs in the infrared (IR) spectral range. The ZnO band gap of ~ 3.40 eV requires optical radiation of frequency of at least ~ $10^{15}$ Hz, which lies in the violet region of visible light spectrum, to excite electrons from valence band edge to conduction band edge or rather from HOMO to LUMO in QDs. To generate LSPRs in the IR spectrum, therefore, requires doping with shallow impurities.

Doped QDs can be synthesized in many ways, but colloidal synthesis has emerged as the main synthetic route among the wet chemical synthesis techniques. Advances in colloidal synthesis make it possible now to grow monodisperse QDs in colloids with sub-nm uniformity. Such a stable, uniform dispersion is achieved with surface surfactants that prevent the QDs from aggregating. In addition, the chemical control over dopant type, concentration, and distribution in the QDs by colloidal synthesis technique opens many possibilities in the design of LSPR – active QDs.[19,20] Such a wide range of possibilities warrants computer experiments to tailor the synthesis protocols.

While it is possible to synthesize QDs as small as 2 nm, it is QDs in the range of 3 – 5 nm that are more easily accessible experimentally. In this work, we performed computational studies of QDs of passivated ZnO, ZnGaO, and ZnAlO of approximately 3 nm in diameter consisting of 1580 atoms, which is a sizable number for conventional density functional theory (DFT) calculations that use diagonalization approach to determine Kohn – Sham eigenvalues and orbitals. We have previously computed ZnGaO of 1.4 nm using plane – wave basis, but the plane - wave basis is computationally too expensive to perform DFT calculations of such a magnitude involving 1580 atoms.[21] In this work, we are able to compute such a QD using a numerical atomic basis. As such, it sheds light on LSPR properties of quasi-spherical QDs with facets, taking geometric and electronic structure of the QD, but also shows the capabilities of current high performance computing clusters that consist of nodes with large random-access memories (RAM).[22] In the following sections, we will first present the computational details in section II, followed by results and discussion in section III, and conclusions in section IV.

II. Computational Details

It is well known that the conventional density functional theory (DFT) calculation in the local density approximation (LDA) or generalized gradient approximation (GGA) for the exchange correlation functional underestimates the band gap of materials.[23] Therefore, the band gap must be corrected to study optical properties. We performed the DFT calculations with Hubbard correction (U) as a parameter,[24] as implemented in Spanish Initiative for Electronic Simulations with Thousands of Atoms (SIESTA) code.[25] As the name suggests, SIESTA is well suited to study a large system such as the 3.0 nm – diameter QD in this work consisting of 1580 atoms. It uses numerical atomic orbitals as basis functions to solve the Kohn – Sham equations.[25] The 3d and 4s orbitals of Zn and the 4p orbital of O are taken as valence electrons. The rest of the electrons are replaced by norm – conserving pseudopotentials, which were obtained from the Cornell University pseudopotential repository.[26] The wurtzite unit cell for bulk ZnO was obtained from materialsprojects.org[27] and was re-optimized with Hubbard U corrections and the Pedrew, Burke, Ernzerhof (PBE) approximation for the exchange – correlation functional.[28] A spherical supercell for the quantum dot (QD) was cut out of the bulk crystal by enlarging the re-optimized unit cell. Such a supercell comprises of triply bonded, doubly bonded, and singly bonded dangling bonds at the facets of the QD with different orientations. The singly bonded dangling bonds were removed as they are unstable energetically. The removal of the singly bonded dangling bonds as well as the facets with different orientations result in a quasi-spherical quantum dot. The remaining dangling bonds were passivated with two kinds of pseudo-hydrogens: the Zn atoms were passivated with pseudo-hydrogen of charge 1.5e and the oxygen atoms with pseudo-hydrogen of 0.5e.[29] The resulting structure is given in Fig. 1. The passivated structure was then surrounded with a vacuum region of 10 Å to create a periodic structure even though the vacuum does not play any role in the computation based on the localized orbitals of SIESTA as it does in codes with plain wave basis.

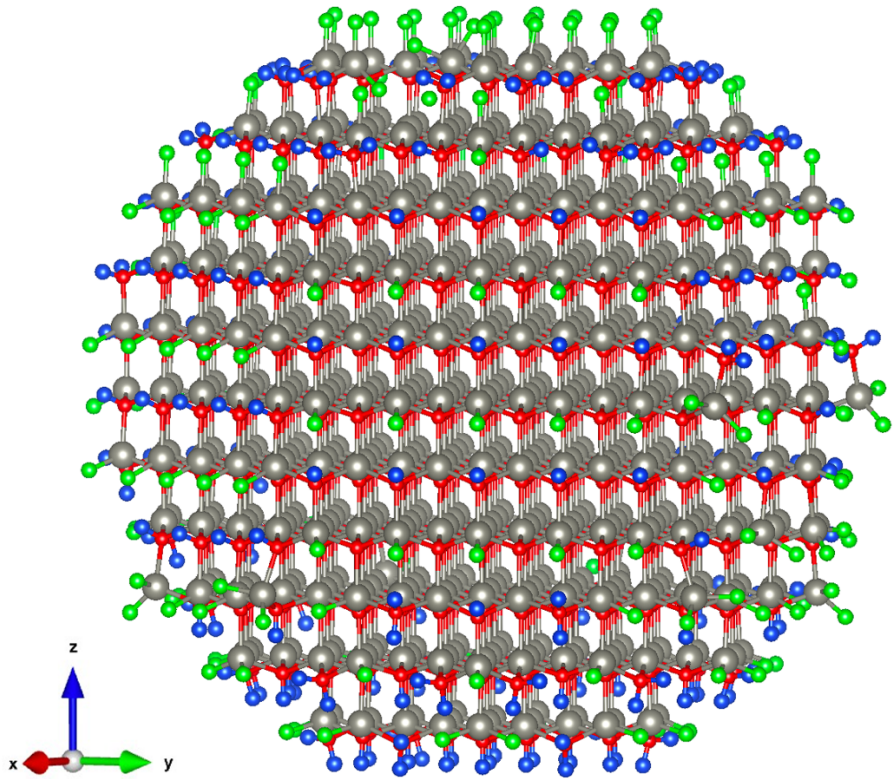

Fig.1. ZnO QD of 3.0 nm in diameter consisting of 1580 atoms. Color scheme: gray – Zn, red – O, green – pseudo H bonded to Zn, blue – pseudo H bonded to oxygen. The size of the atoms is not to scale, but rather used to show the atoms clearly.

The geometrical optimizations were computed at gamma point only for such a large quantum dot first to obtain the ground state. The electronic self-consistent field (SCF) calculations were done using the Broyden charge mixing scheme until an energy difference tolerance of 0.1 meV was obtained. The SCF calculations were repeated for different ionic configurations until a force tolerance of 0.04 eV/Å [30] Is achieved with FIRE molecular dynamics at electronic temperature of 1 K. After the geometric optimization is completed, first order time dependent perturbation theory was applied; the dipolar transition matrix between ground state eigenfunctions was calculated to obtain the imaginary part of the dielectric function.[31] Two kinds of optical meshes,1x1x1, the gamma point, and 9x9x9 were used to integrate in the Brillouin zone. The 9x9x9 integrations were done to see the effect of increased k points. The real part of the dielectric function is then computed with post-processing script supplied by SIESTA with Kramers – Kronig transformation from the imaginary component. Finally, the absorption properties as well as other optical properties are computed from the real and imaginary parts using a second post-processing script supplied by SIESTA.

The optical calculations were mostly performed for [001] electric polarization. To compare the effect of polarization, however, we performed a limited set of calculations for polarization along [100]. In all the optical calculations, only excitations from shallow impurity levels to low lying states above the LUMO were considered by restricting the photon energy to less than 3.0 eV, which is less than the band gap to avoid excitations from HOMO states. We note the energy range of 3 eV extends beyond the discrete levels into the band portion of the eigen spectra, as seen in Fig. 2. The effect of this extension will be used in the interpretation of the absorption spectra later. The focus of the work was on the 3.0 nm QDs, but to study the size dependence of LSPRs, we investigated 2.0 nm QDs as well. We attempted to study 4.0 nm QDs but found out the convergence of the geometric optimization was too slow.

According to Lounis et al, the spatial distribution of dopants affects the LSPR response of metal oxide nanocrystals.[32] They found out frequency – dependent plasmon damping due to impurity scattering is suppressed when the dopants are incorporated near the surface. In colloidal synthesis, the spatial distribution of dopants can be controlled through kinetics of dopant precursors, resulting in either uniform or surface segregated incorporation. Dopant precursors with slow decomposition rate result in near surface placement of the dopants.[32] The question then arises as to where to incorporate few dopants in QD with many facets. One could calculate the formation energies on different sites of the facets: centers, edges, and vertices, but there are as many choices as there are facets and this makes such calculations computationally cumbersome. Instead, we assumed the potential energy at the center of the facet is lower than at the edges and vertices. For few – dopant calculations of this work, the number of facets is more than the number of dopants, which raises another question of which facets to choose. We chose them based on the nature of the impurity state of a shallow donor and electron – electron interaction.

The localization radius of the excess electron of a shallow donor is much larger than the lattice constant. On the average, the excess electron is located far from the center and is weakly bound to it. This means that the atomic structure of the impurity center has little influence on the excess electron. In this picture, the electron can be thought of moving in a central potential of the positively charged center scaled by the dielectric permittivity of the host lattice.[33] Therefore, there is freedom to place the dopants on facets that are computationally easy to handle subject to the condition that electron – electron Coulomb repulsive force among the few electrons would distribute them as far apart from each other in the tiny nanoscale volume of the 3.0 nm QD. The electron – electron interaction in ZnO QD, with dielectric permittivity of approximately 10.4, however, is assumed substantially screened by the dielectric medium to justify the DFT calculation.

III. Results
The size of the HOMO – LUMO gap of small QDs is size dependent. Therefore, to validate our DFT + U calculation with SIESTA code, we calculated the band gap of bulk ZnO using U = 14.0 for Zn d orbitals and U = 9.0 for O p orbitals, which yielded a gap of 3.35 eV that is in good agreement with experimental band gap of ~ 3.40 eV.[34] To validate further, we plotted the density of states (DOS) and partial DOS in Fig. 2. The DOS agrees well with our previous results using a VASP code,[21] and the DFT calculation that uses HSE06 hybrid functional for the exchange – correlation functional.[35] The PDOS make up is also consistent with results in the literature, that the bands below the valence band maximum consist of s and p orbitals of oxygen, followed by the spikey DOS that consists mainly of d orbitals farther below. While these results give us confidence in the DFT + U computations, for our subsequent calculations involving shallow impurities and optical transitions to low lying states near LUMO of the QDs, it is the structure of the conduction band that is relevant. Here again, the DOS agrees well with our previous calculation and to that of the HSE06 calculation. What is even more relevant to subsequent interpretation of our results is the PDOS of the low-lying states that determines the kind of orbitals available for optical transitions from the shallow impurity levels subject to selection rules of the electric dipole transition matrix of the first – order time dependent perturbation theory. We note here for the large supercells of doped ZnO, only gamma point calculations are performed since the Brillouin zone corresponding to the large real pace volume of the QD is small, and the concept of DOS fails as it requires a dense mesh of k points in the Brillouin zone. So, the results of the bulk unit cell will be used as approximations later during the discussion of the optical transitions from the shallow impurity levels to the low-lying states above the LUMO of the QD.

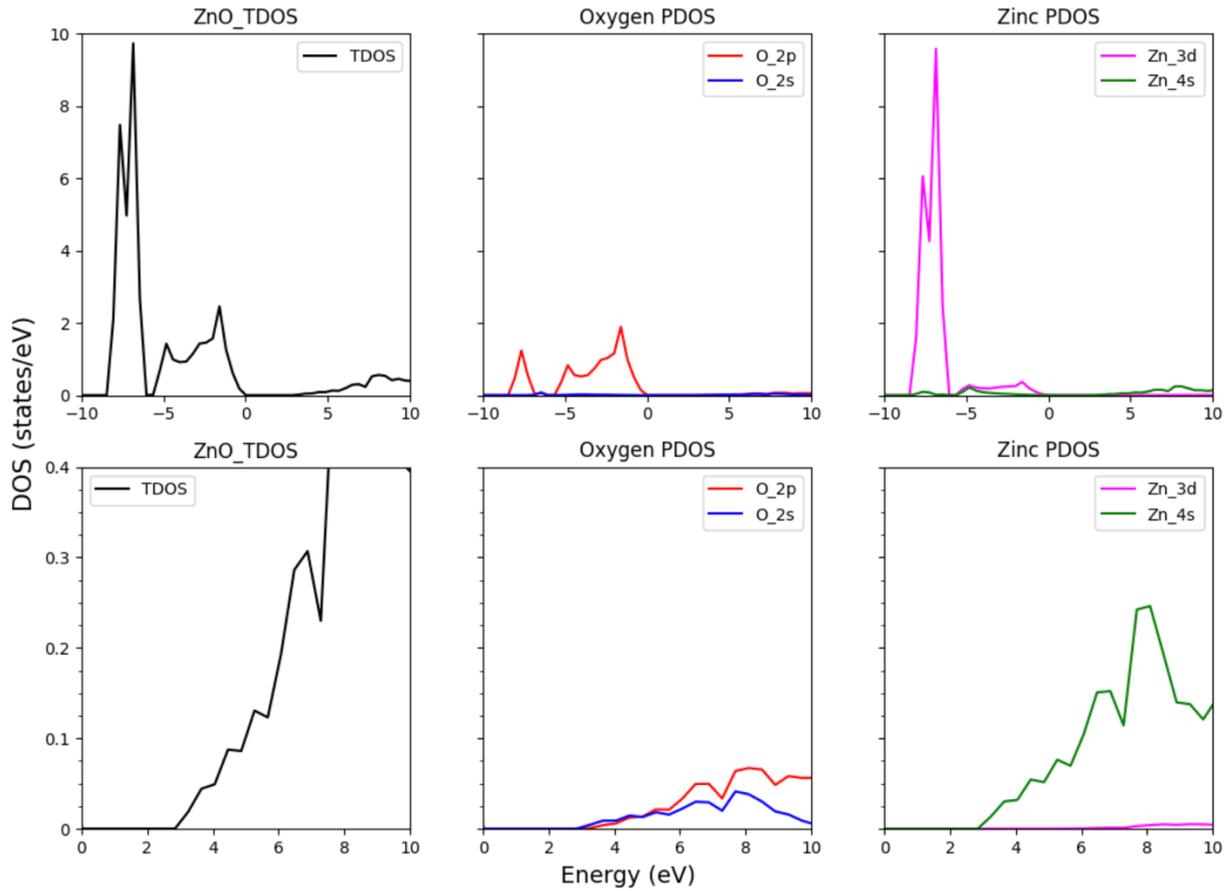

Fig. 2. Total density of states (TDOS) and partial density of states (PDOS). The DOS scale in lower panel is blown up to show the PDOS near the LUMO.

To find out the nature of the eigenenergy spectra of the low-lying states and the location of the impurity level, we computed eigenvalues for one dopant at representative points: the center of the QD, the center of facet, and the center of a facet one layer above a polar facet that terminates in O ions. In Fig. 3, we plotted the eigen spectra for the three cases and juxtaposed them to the undoped eigen spectra to see where the impurity levels are. The bandgap for the 3.0 nm QD is larger than the bulk, as expected. The shallowness of the impurity levels, and the discreteness of the spectra near HOMO and LUMO is clearly seen.

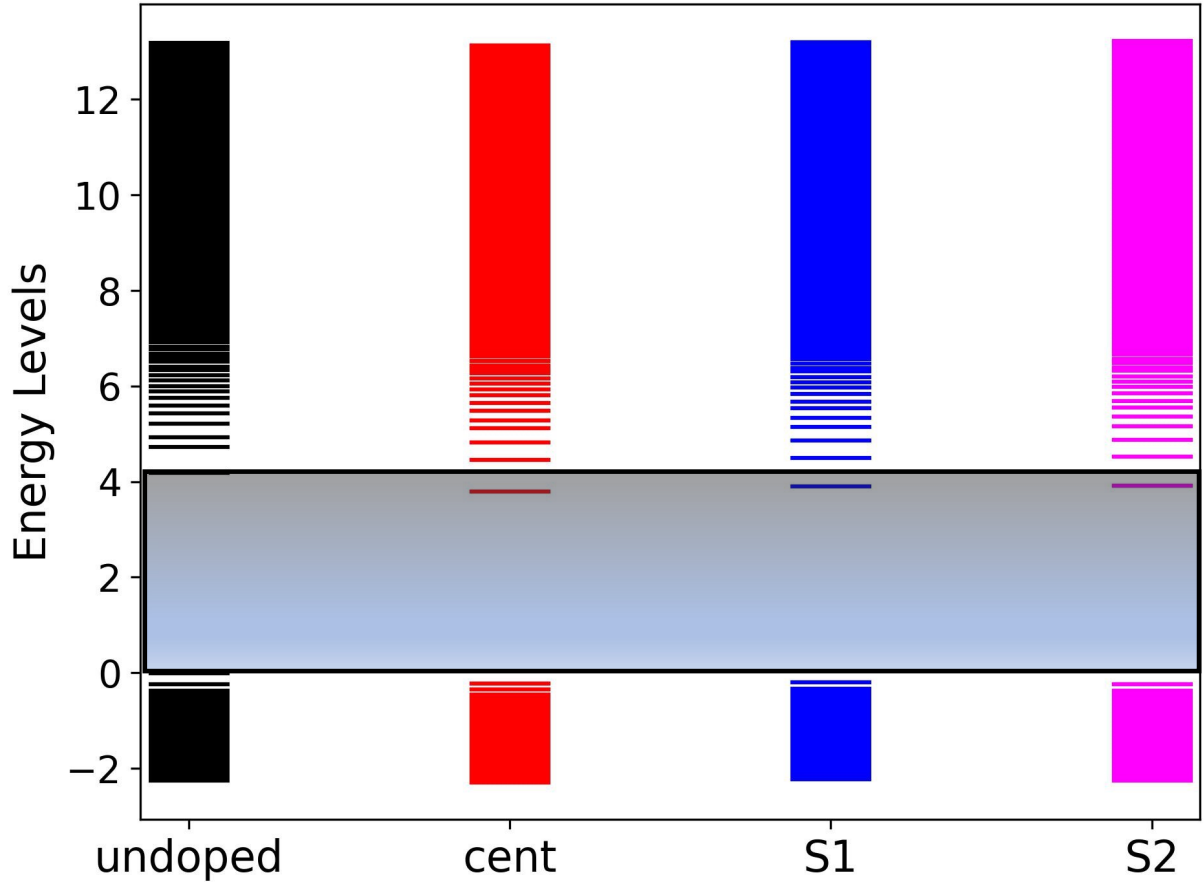

Fig. 3. Eigenvalues spectra of undoped ZnO, singly doped ZnO with Ga at center (cent), at center of Zn – terminated polar facet (S1), and at center of Zn facet above O – terminated polar facet (S2). The undoped spectra is juxtaposed for comparison. The shaded area is the gap.

**A. Absorption spectra**

The intent of this work is to find out if the absorption spectra of few electrons have resonant absorptions. The resonant absorptions are expected to excite the hydrogenic electrons from shallow impurity levels to the low-lying states above the LUMO and cause the electrons to oscillate at the resonant frequency of the oscillating electric field. Furthermore, since the electrons are few, the Coulomb interaction among them will spread them out spatially, subject to local potential variations, but also correlate their oscillation resulting in plasmon resonances. To gain insight into how such plasmon resonances could arise for few electrons and how they are related to electronic and geometric structure of the QDs, polarization of the electric field, and type of dopants, we computed the absorption coefficients from the dielectric function.

As stated in the introduction, we chose the ZnO QD because of its simple stoichiometry. However, its wurtzite phase at room temperature is anisotropic and the dopant distribution must be chosen accordingly. In addition, the finiteness of the QD, with local variations around the surface as seen in Fig. 1, complicates the choice of the distribution. One could apply a statistical method to distribute the dopants based on site energies but determining the site energies quantum mechanically for such a large system involving 1580 atoms of 5 different species (Zn, O, dopant, 2 kinds of pseudo hydrogens) is computationally intensive. To get insight into LSPR evolution, we distributed the dopants at or around centers of facets: two of the facets defined by crystallographic directions of the wurtzite crystal, [100] in the basal plane (or **a**), and [001] (or **c**), and the third kind defined by the Cartesian direction [010] instead of the **b** direction at 120º with respect to **a**. The polarity of the paired facets (001) and (00$\bar{1}$) means that one of the Zn facets is chosen one layer above the O terminated facet.

The focus of this work is on six – dopants which can be distributed uniformly along the chosen directions. To help us analyze the excitations of the six-dopant configuration, we investigated the absorption spectra of one-electron, two-electrons, and four – electrons first to see if there is some kind of trend for spectral lines. The odd configurations of three and five were skipped as they do not distribute uniformly along the chosen directions.

### B. One - electron absorption spectra

A pertinent issue of the absorption spectra is how to distinguish between single electron transitions and collective electron transitions of plasmon resonances, and whether such resonances arise due to intraband or interband transitions. As seen in Fig. 2, the eigenvalues spectra for low lying states are discrete and not band-like near the LUMO. Therefore, the concept of intraband or interband transitions does not hold. In addition, there is no analytical expression for the dielectric function of few electrons, as that for highly doped QDs in the Drude model, as well as the electronic polarizability, the singularity of which is used to locate the resonance frequency. In light of this shortcoming, we resorted to identify first the one electron spectra with one dopant to help us distinguish between the single electron spectral lines and few – electron spectral lines. We assume if few one electron excitations are involved, the photon energy involved is the same for all the one – electron excitations, therefore, only the magnitude of the spectral line will increase. There could be slight difference in the photon energy due to local potential variation that could lead to broadening of the spectral lines. A Gaussian optical broadening parameter of 0.02 eV is applied to the spectral data to smooth out spikey spectral lines.

Fig. 4. shows the absorption spectra of one - electron excitation for three different sites: center of the QD, and centers of the polar surfaces. Since only one electron is involved, these spectra can only be due to one electron excitation. We computed the spectra for gamma point, 3x3x3, 5x5x5, 7x7x7, and 9x9x9 optical meshes in the Brillouin zone, but present results of the gamma point and the 9x9x9 mesh. The results of the rest of the meshes are similar to that of the 9x9x9 mesh and are given in supporting information. The non-dependence of the absorption spectra on the number of k points shows that it is the energy difference between the impurity level and specific energy of the low-lying state with specific k point that is involved, analogous to photoelectric effect. Moreover, the result validates that the gamma point-only calculation for a supercell is sufficient to capture the absorption spectra.

The results are also consistent with our earlier assertion that hydrogenic wavefunctions do not depend on the local structure of the dopant ion. In all cases, there are two peaks, one lower in magnitude than the other. The photon energies corresponding to the peaks for the surface dopants vary very slightly from that at the center, but the magnitudes of the peaks are almost the same. The two spectral lines could be explained by electric dipole selection rules for a hydrogenic electron in spherically symmetric potential. This is an approximation since the space around the dopant ions on the scale of the localization radius of the excess electron is not strictly isotropic at the surface. Assuming the excess electron of the dopant is in s state, the longer spectral line can be assigned to the allowed transitions to p state. We conjecture that the shorter spectral line is due to transitions from the impurity level to s orbitals hybridized with d orbitals, based on the structure of partial orbital density of states in Fig. 2. Irrespective of the interpretation of the spectral lines, however, the spectral lines can be identified as signatures of one electron excitations and will be used to distinguish spectral lines when few electrons are excited. We note also there are two distinct parts to the overall spectra; there is the portion of two spectral lines associated with the discrete part of the energy spectra followed by somewhat exponentially decreasing part associated with the band portion of the energy spectra in Fig. 2. The energy range used in the calculation, that is 3 eV, includes both the discrete and band portion of the spectra as seen in Fig. 2. The spectrum for the discrete part suggests that the numerical atomic basis is the natural basis to describe the absorption of the QDs unlike the plane wave basis used in our earlier work which is more appropriate for the band portion seen in Fig 4.[21]

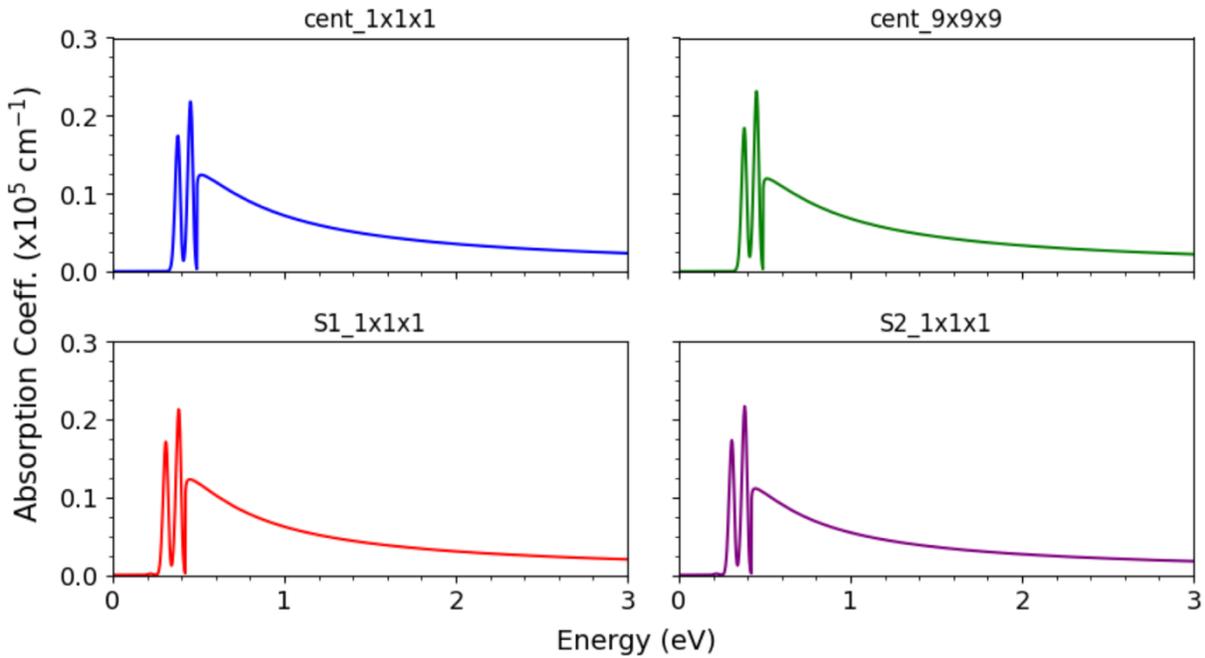

Fig. 4. One-electron absorption spectra for single dopant at center (cent), S1 (001), and S2 (00$\bar{1}$). S2 is the Zn facet above the O – terminated polar facet. The Brillouin zone integration is performed at gamma point (1x1x1) and 9x9x9 k-points.

C. **Two-electron absorption spectra**

We placed a pair of dopants on polar facets of (001) and (00$\bar{1}$), opposite facets (100) and ($\bar{1}$00), (010) and (0$\bar{1}$0), and adjacent facets (001) and ($\bar{1}$00), (00$\bar{1}$) and $\bar{1}$00), (001) and (0$\bar{1}$0). (00$\bar{1}$) and (0$\bar{1}$0) were chosen to pick the Zn atoms on the outermost facets. All the chosen facets have the two spectral lines like those in one - electron excitations, albeit in varied form from each other with different magnitudes, and corresponding photon energies. This shows the distribution of the dopants matters unlike the one dopant case above or the jellium model applied for highly doped system.[15] Only the pair on (001) and (00$\bar{1}$) surfaces have sizable spectral lines next to the two spikey spectral lines of the one – electron excitations. We attribute these to collective or paired excitations. It is tempting to ask then why the other pairs do not manifest such sizable secondary spectral lines. This is due to the polarization of the electric field which is in the [001] direction, as a result the polar charges on the (001) and (00$\bar{1}$) form a stronger dipole moment. The stronger dipole moment causes correlated transitions. The photon energy for the paired excitation, however, is larger than those for the two one – electron excitations. Moreover, the magnitude of the peaks of one – electron excitations is much larger as well. This implies the one – electron excitations are more likely than the paired excitation of two electrons.

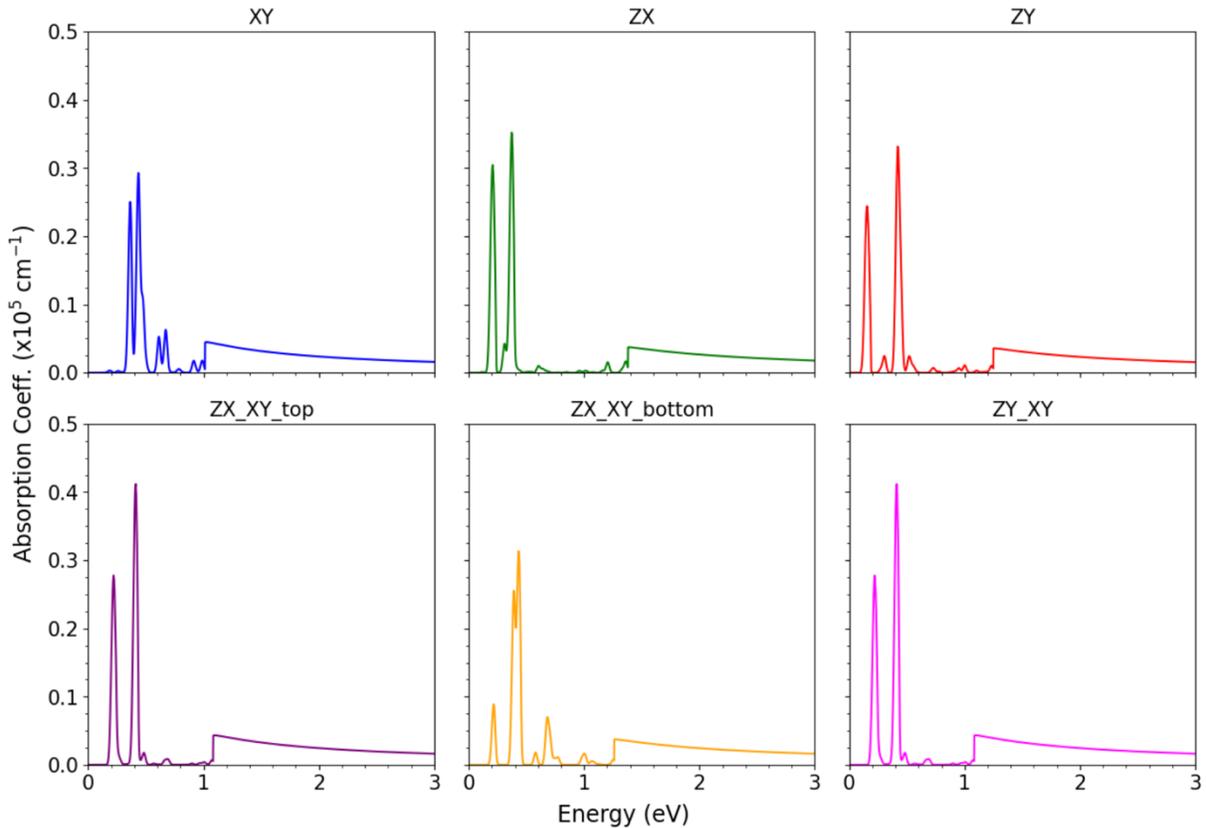

**Fig.5.** Two-electron absorption spectra. The xy surface is defined by x and y axes with one dopant on (001) and second one on (00$\bar{1}$) facets. Similarly, zx and zy consist of two dopants on opposite facets defined by corresponding axes. Two dopants with one dopant on zx and a second one on xy surface are denoted by zx_xy. Similarly, for zy_xy.

**D. Four-electron absorption spectra**

Two kinds of facet configurations for dopants are chosen. In both configurations, the pair on (001) and (00$\bar{1}$) polar facets are retained. The other two dopants were placed on (100) and ($\bar{1}$00) facets in the first configuration and on (010) and (0$\bar{1}$0) in the second case. Each of the two types of configurations were repeated on inner layers of Zn and the absorption spectra for each type are shown in Fig. 6. The dopant ions get closer to each other for inner configurations. Such configurations will show two effects: first, they will show the effect of outermost facet configuration relative to the inner configurations and second, and the effect of interionic distance between the dopant ions on the absorption spectra of the electrons.

It is stated earlier that the hydrogenic electrons are so far away from the impurity center that they are not affected by the impurity center. This is true for isolated one dopant in a homogeneous isotropic crystal structure. For few electrons in volume of a 3.0 nm nanostructure that is anisotropic, however, the assertion of the one-electron does not hold. The four – dopants configuration is not spatially uniform; one of the pair of facets, either (100) and ($\bar{1}$00) or (010) and (0$\bar{1}$0) is undoped when the other is, creating potential variation in the QD. As a result, the absorption spectra show more structure.

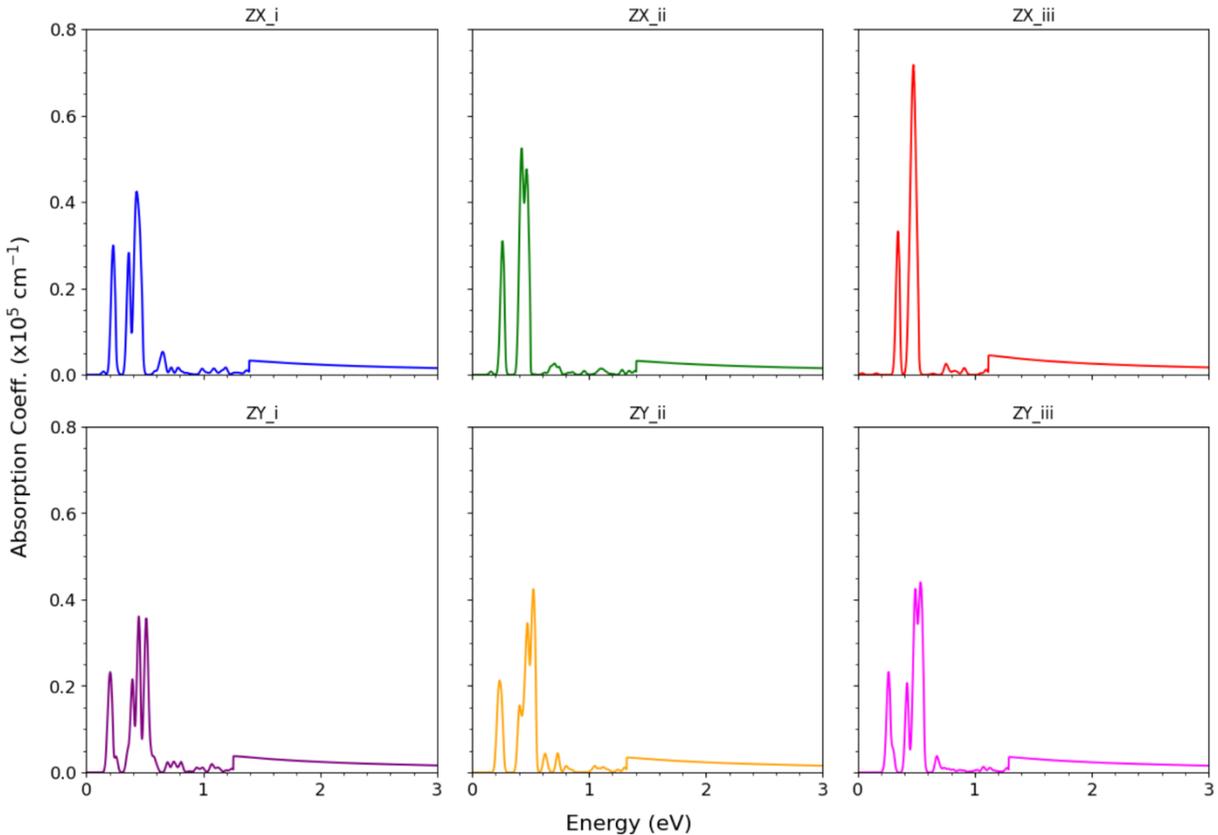

**Fig. 6.** Four - electrons absorption spectra. Two dopants are placed on xy polar facets in all cases. Outermost configuration is denoted by i, the middle one by ii, and innermost configuration by iii. In upper panel, the other two dopants are placed on zx facets while on the bottom panel they are placed on zy facets.

Fig. 6 contains two sets of absorption spectra one for {100} facets and a second one for {010}. Each set consists of outermost and two inner facets. In {100} case, the two single spectral lines of one – electron excitations have broadened and combined with each other significantly. The secondary spectral line on the lower side of the one – electron excitations is due to many-electron excitations. As the facets move inward, the interionic distance gets shorter and the interaction energy increases, and the broadening of the spectral lines for the one-electron excitations increases, culminating in one single-electron spectral line. The secondary spectral line for the lower side photon energy remains the same in all three cases. The {010} configurations show similar general trends, i. e., the spectral lines broaden as the ionic distance decreases for the inner configurations. The spectral line corresponding to the side photon energies corresponding to many – electrons collective excitation is the same for all the configurations. The resilience of the many - electrons spectral lines despite the variation of the potential for the configurations implies the few electrons collective oscillations will not be easily perturbed once they are set. Likewise, in the lower panel also, we see the spectral lines for collective excitations are the same for all the configurations despite increased Coulomb interactions for inner configurations. The spectral line due to single electron excitations has three spikes in contrast to two in the case of {100} due to stronger potential energy variation on the facet. Overall, the results show the distribution of the dopants, be it along facets in different crystallographic directions or radially, has a strong impact on the structure of the absorption spectra.

**E. Six-electron absorption spectra**
The six dopants, with a pair of dopants on opposite facets along each Cartesian coordinate, were placed at outermost facets and two more inner ones on Zn layers below the surface. They were studied to compare absorption spectra between inner configurations and the outmost configuration. The results are shown in Fig. 7. The spectra can be identified using the spectra of the one – electron excitation in Fig. 4. The spectral line of the outermost configuration with maximum peak is similar to the one-electron spectral line of the one-dopant configuration. However, the second adjacent spectral line seen in one – dopant configuration disappeared. A plausible explanation for the disappearance could be due to strong electron – ion interaction between spatially uniform distribution of the ions resulting in cancelling the effects of each other on the spherically symmetric s - orbitals. Only the spectral line due to the p – orbital is retained. This central spectral line, with maximum peak, due to one – electron excitations is resilient and is observed in all the spectra studied in this work. The secondary spectral lines at lower energy levels are consistent with the argument that many – electron collective excitations are favorable to individual one – electron excitations because they require less energy.[35,36] The peaks of the lower side spectral lines are almost four times those of the upper side spectral lines, on the other side of the single electron spectral line. The appearance of the secondary few – electron spectral line with sizeable magnitude at lower energy implies the few electrons correlated excitations are just as likely as the one – electron excitations. Which one of them results in plasmon resonance will be shown in section IV of plasmon resonance below.

The spectra from the two inner configurations clearly show that the spectral lines are broadened due to Coulomb interactions and finally merge for the innermost one, showing the effect of dopant distribution. The broadening for the six – dopant case is relatively more than that for the four – dopant spectra underscoring the significance of interionic distance for getting spectral lines with narrower line widths. In addition to broadening, it is worth noting that the peak of the one electron spectral line is the same for the outermost facet and the middle one but increases for the innermost configuration. This raises the question of the possibility of one electron excitations increasing as the Coulomb interaction increases which may be due to the positive ion – ion interaction raising the impurity levels closer to the LUMO level. We assume the electron – ion distance on the average is farther than the ion – ion distance resulting on the overall increase of the positive energy.

The results so far are for gamma – point calculation in Brillouin zone. The eigen spectra near the LUMO are mostly discrete and become band – like as the energy increases. The absorption spectra correspond to excitation from the impurity levels to the low – lying discrete states for photon energies of up to 3 eV considered in this work. To see the effect of more k points in the Brillouin zone, the optical mesh is increased to 9x9x9. We considered only the outermost configuration because it resulted in well - defined spectral lines. The results are given in the lower panel of Fig. 7. Comparing the 9x9x9 mesh to the 1x1x1 (gamma) mesh result, the peak of the one electron spectral line increased, so did one of the peaks of the collective spectral lines on lower side of the one electron spectral line. The photon energies, corresponding to the peaks, remained the same. A plausible reason for this could be inferred from the PDOS. One of the two orbitals, the p orbital with higher energy, is more available as the number of k points is increased, hence, the increased magnitude of the spectral line relative to one another.

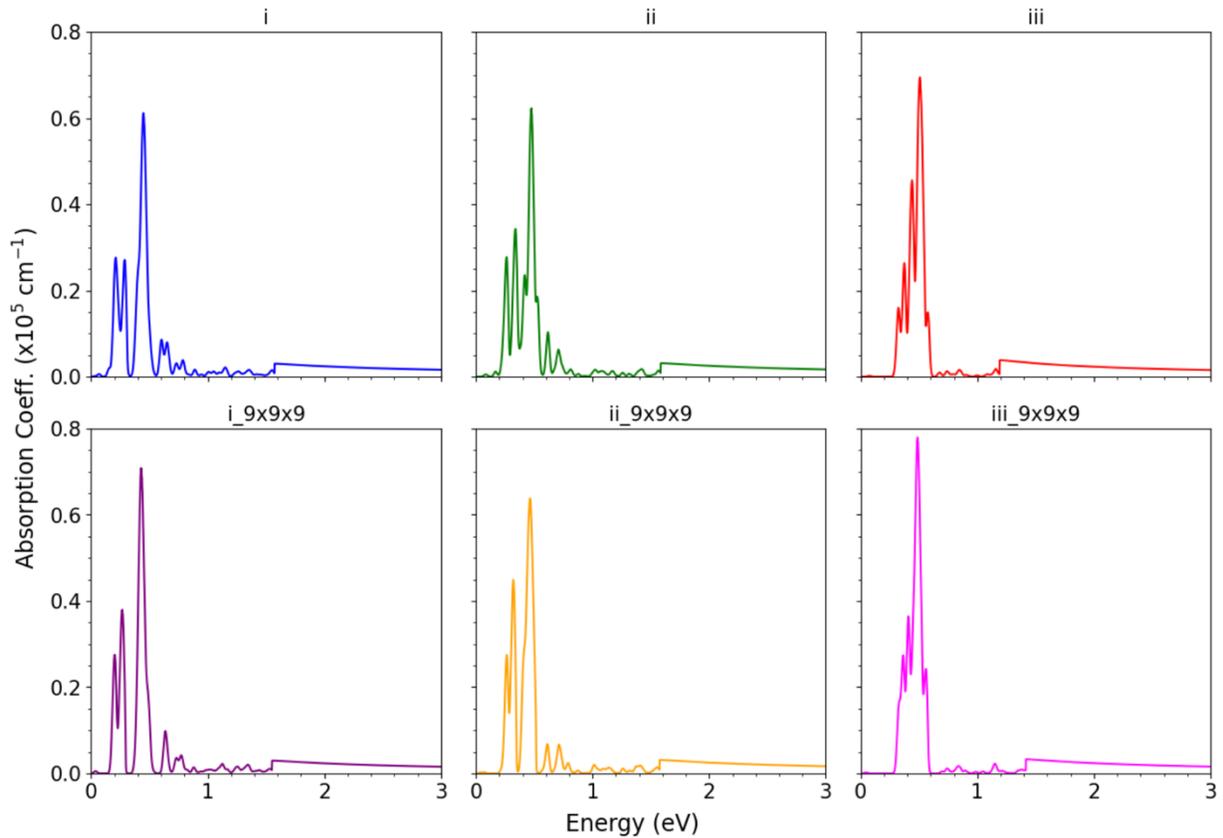

Fig. 7. Six-electron absorption spectra. The three configurations at outermost, middle, and innermost facets are denoted by i, ii, iii and are computed at gamma point. The corresponding 9x9x9 k-points results are given in lower panel.

**F. Electric field polarization**

It has been reported that the orientation of facets of QDs in a colloid can be controlled through small passivating ligands.[37,38] Assuming the surface orientation can be controlled, then an electric field can be turned on in different directions. Our results so far are due to electric field in **c** direction of the wurtzite crystal. To see the effect of orientation of electric field, we turned on the electric field in direction perpendicular to the c axis, along the other crystallographic direction [100]. As seen in Fig. 8, the choice of the electric field has a significant effect on the absorption spectra. The s and p conduction electrons will respond differently. The p electrons will reorient themselves in the direction of the electric field as they are being excited and will oscillate in phase with the electric field. On the other hand, the charge distribution of the s electrons will be distorted by the electric field resulting in asymmetric distribution. Thus, the electrons oscillating in different directions will sample different local potentials resulting in different absorption spectra. The double spectral lines of one - electron excitations are still observed but there are slight variations; the p and hybridized s spectral lines are equal in contrast to the p - electron spectral line being slightly larger for the c polarization. On the other hand, the six – dopant case shows significant change between the two spectra due to orientation of the electric field. The spectral lines for the six electrons are at the same energy levels as those for one electron, therefore, are not correlated few electron excitations. The magnitude of the spectral lines has changed due to more electrons being excited at the same energy. The spectral lines have broadened due to Coulomb interactions, in particular for the p spectral line. We attribute this is due to the electrons oscillating along **a** direction seeing a significantly different potential than those along c direction.

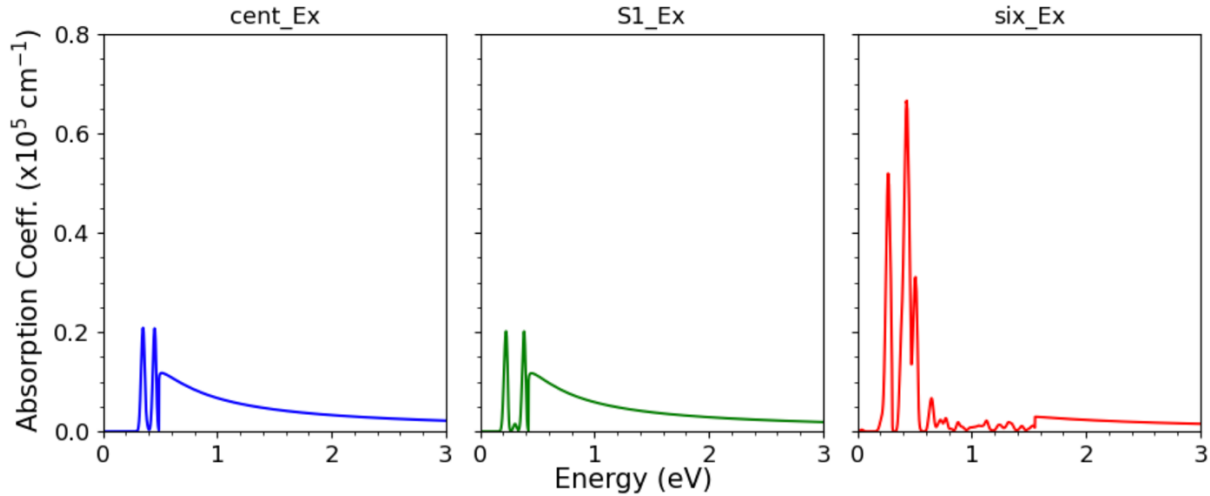

Fig. 8. Absorption spectra with electric field oriented along x – axis or a crystallographic direction in hexagonal basal plane. The first two are for one – electron excitation, and the last one is for six electrons.

## G. Comparison with 2.0 nm QD

We tried to compute ZnGaO QD of 4.0 nm, but the number of atoms increases from 1580 for 3.0 nm to 3840 for 4.0 nm, and the optimization to determine the ground state was too slow despite having nodes that could handle the increased memory requirements for RAM. Instead, we investigated the 2.0 nm QD which has 689 atoms, less than half the number of atoms for 3.0 nm. In Fig. 9, we present results for one dopant at the center and on the surface first for comparison and the six dopants on the outermost facets as before. The results show that the one-dopant spectra do not have the two spectral lines of the 3.0 nm QD due to the increased interaction energy between the electron and ion, and surface scattering. The two lines due to transitions to p and hybridized s orbitals were combined to one spectral line. The relatively small lower side spectral line for the surface case is the remnant of the hybridized s orbitals. The six – electron absorption spectra are similar to the second case (ii) of Fig. 7 for the 3.0 nm QD. As we noted earlier, the inner configurations have closer dopant distribution with stronger Coulomb interaction, therefore, in a way not surprising their spectra are similar to that of the 2.0 nm case with close dopant distribution and strong Coulomb interaction.

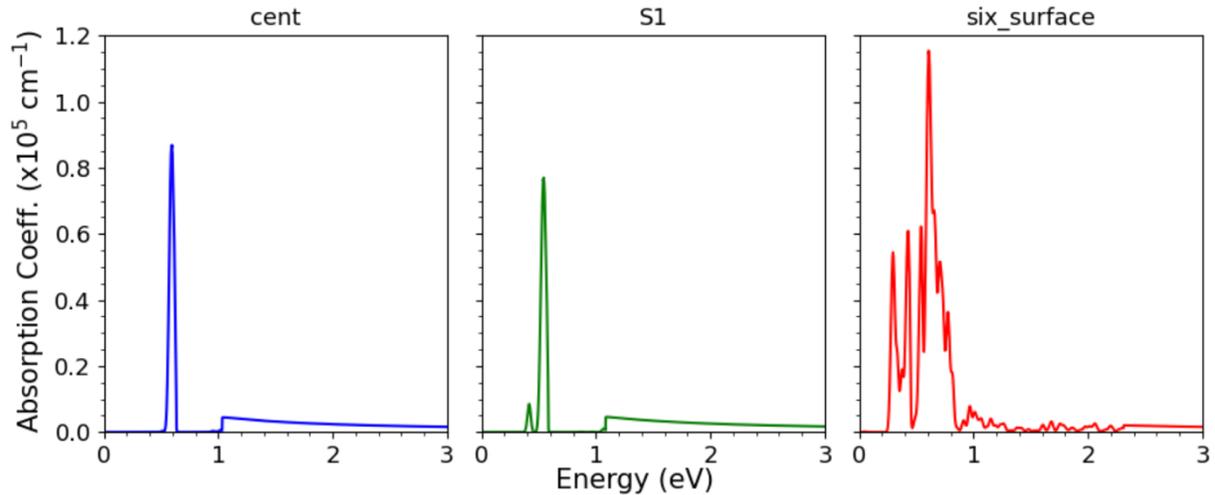

Fig. 9. Absorption spectra for 2.0 nm ZnGaO: for single dopant at center, surface, and six dopants at the outermost facets.

## H. Al as a dopant

Al, Ga, and In atoms belong to the same column in the periodic table and have s and p valence electrons in their outer shells equal to $ns^2$, $np^1$, n=3, 4, 5 for Al, Ga, and In respectively. They are all used to dope ZnO. However, we contrasted only the absorption spectra of the six-dopants ZnAlO and ZnGaO in Fig. 10 because the Cornell pseudopotential repository, whose pseudopotentials used in this work, did not have the SIESTA pseudopotential for In. The absorption spectra for Al and Ga are similar. The broadening of the spectral lines is more for Al due to difference in pseudopotentials and ionic radii. The Ga pseudopotential includes the core localized 3d orbitals which screen off the Coulomb interaction more than the core s an p electrons of Al. As a result, the Al spectral lines are broadened more. In addition, Ga is almost the same **size** as Zn while Al is slightly less, which results in potential variation due to effect of strain of dopant substitution.

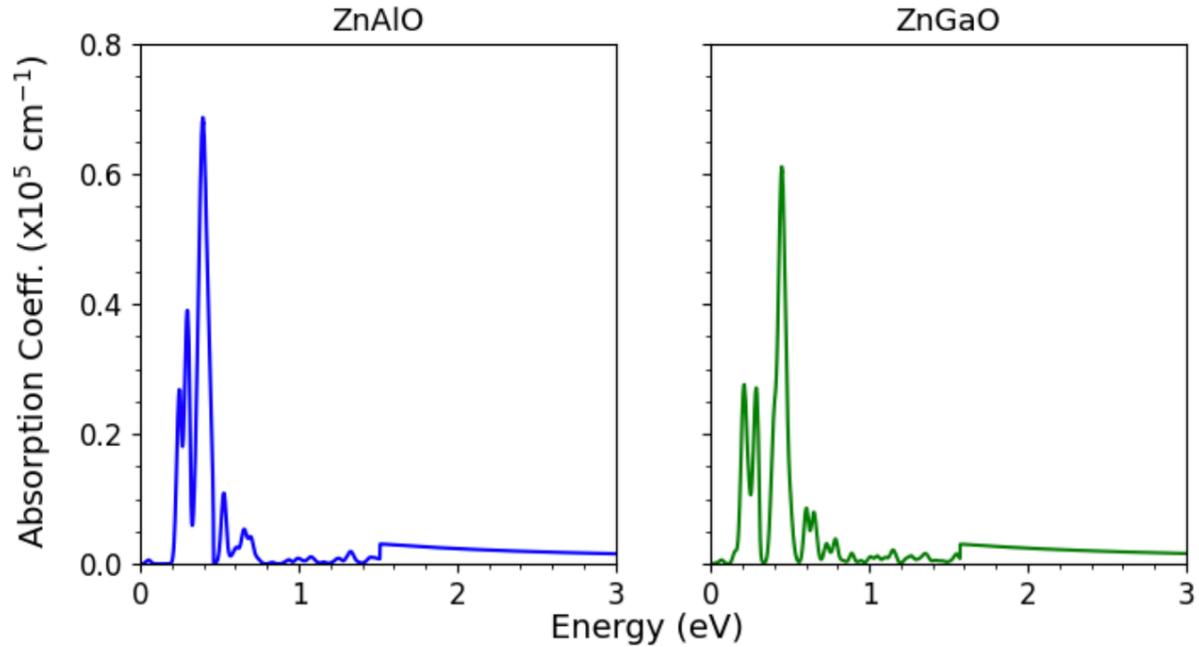

Fig. 10. Absorption spectra of Al vs Ga for six electrons. The dopants are placed on outermost facets.

IV. Plasmonic resonance

The photon energy excites the electrons from the shallow impurity levels to the low-lying states near the LUMO. We see resonance peaks of different magnitudes in the absorption peaks of most of the configurations that were studied. Will these resonant frequencies of absorptions be maintained in the resulting oscillations of the electrons in the low-lying states? In the simplest model of uniform electron gas confined to constant potential well, the oscillating electric field will cause the electrons to oscillate in phase. However, in realistic model of finite QD, the electrons will scatter off the facets. We assume the electron – electron scattering among the few electrons in the lightly doped limit is not as significant as the surface scattering as the electrons will be as far apart from each other due to mutual Coulomb repulsions. For large QDs of 3 nm or more with relatively large facets, compared to the 1.4 nm QD studied with VASP [21], the electrons may adjust their oscillations locally while interacting with each other and could continue to oscillate in phase. To see if such a possibility exists, we plotted the real parts of the dielectric function vs photon energy as shown in Fig. 11 upper panel for the six dopants case on the surface for gamma point and 9x9x9 optical mesh. The real part of the complex dielectric function is a signature of LSPRs when it is negative.[17] In Fig. 11, we find for two photon energies, ~ .25 eV and ~.41 eV, the real part of the dielectric function is negative, and the corresponding resonant absorptions with almost same energies are seen in the lower panel of Fig 11. The tiny variations in the two sets of energies could be attributed to scattering. The results clearly show that one can tune the absorption frequency to that of the absorption energy of ~.41 eV to get a significant LSPR. To see if there are LSPRs associated with other k points, we increased the optical mesh to 9x9x9 as stated above. We see that the plots for real dielectric function are the same for both the gamma point and 9x9x9 mesh. This shows that the LSPRs are determined by the absorption or excitation energy only. The interpretations of intra-band transitions in the highly doped limit as signatures of LSPRs may not be applicable to the discrete energy spectrum of the low-lying states of few electrons.

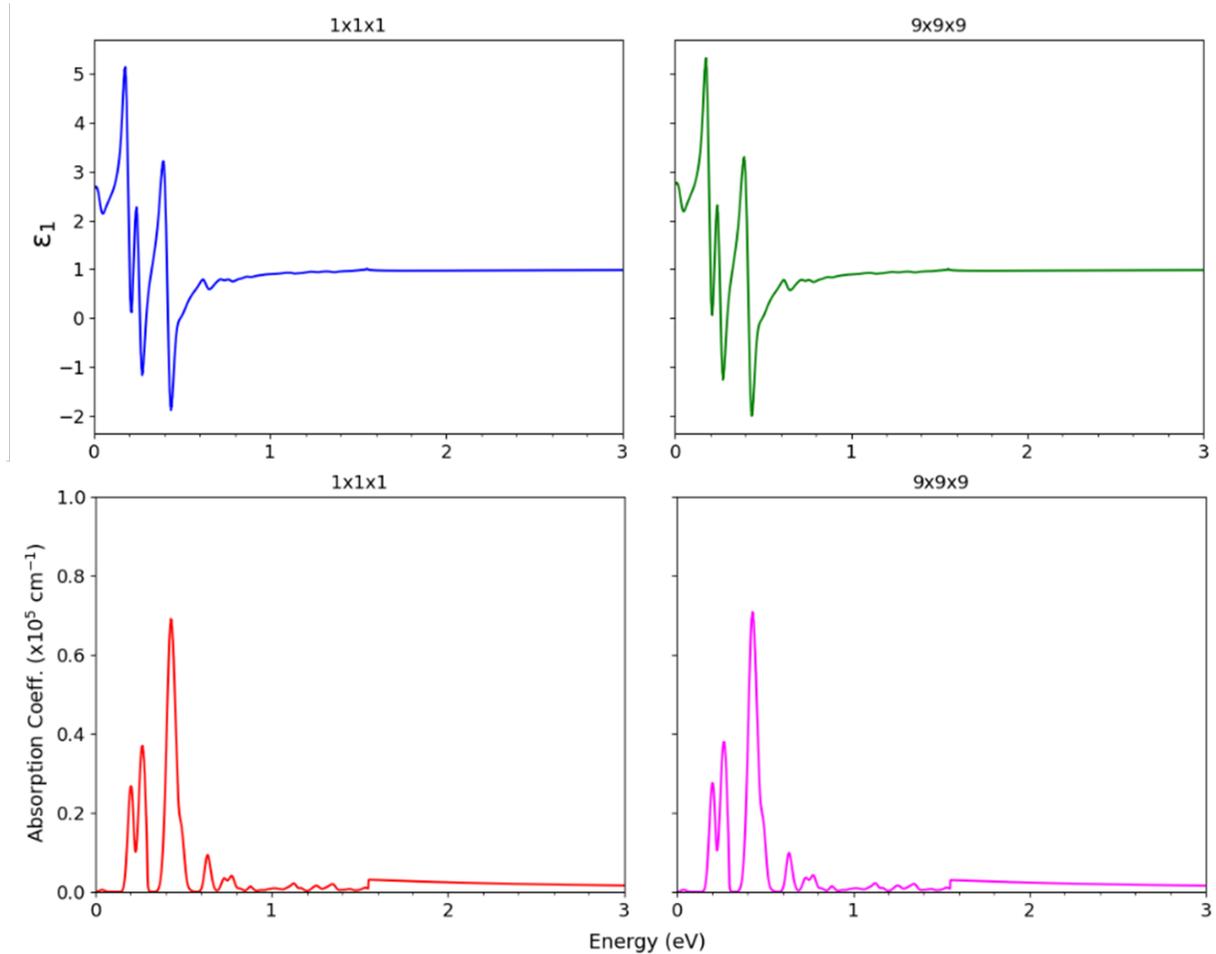

Fig. 11. Upper panel, real part of the dielectric function, and lower panel, absorption spectra for 6 dopants on the outermost surface of the ZnGaO QD of 3.0 nm.

V. Conclusion

We studied in detail absorption spectra of a sizeable QD of ZnO of 3.0 nm doped with Ga mainly, and Al for comparison. We found uniform distribution of dopants on facets around outermost surface results in spectral lines with narrower line widths. This is a welcome characteristic since one of the challenges encountered in doping of nanocrystals has been the diffusion of dopants to the surface instead of being uniformly distributed throughout the volume and the phenomenon has been termed self – purification.[39] The absorption spectral lines can be correlated with plots of the real parts of the dielectric function to determine plasmon resonances as shown in Fig. 11 above. For the uniform distribution around the outermost surface, the plasmonic modes consist of p orbitals mainly, assuming the impurity – level hydrogenic electrons consist of s orbitals. Such correlations between absorption spectral lines and the negative portion of the real part of the dielectric function, coupled with PDOS, can be employed to study the effect of complex factors on plasmon resonances such as surface scattering. For instance, as expected, comparison of the 3.0 nm results to 2.0 nm shows much more broadened spectral lines for the 2.0 nm QD.

We have also shown that such structural studies can be used to filter out an optimal dopant for plasmon resonance. While Al is commonly used to dope ZnO experimentally, in this work, Ga is found to be a better dopant than Al, with narrower spectral line widths because of the localized 3d electrons screening off Coulomb interactions as well as the size of Ga being closer to Zn than that of Al, reducing the potential energy variation due to strain. The result underscores the relevance of electronic structure to design or study LSPR – active doped QDs. For instance, the PDOS of the conduction band, even though it is determined for bulk unit cell in this work, could approximate the electronic structure of large supercells such as the 3.0 nm one used in this work and can be used to figure out the orbitals involved in electron transitions. Unlike bulk, however, the discrete nature of the energy spectrum near the LUMO, as shown in Fig. 3, must be taken into account while interpreting results for plasmonics of few electrons. For instance, classification of transitions into intraband and interband is not applicable to discrete energy spectra of low-lying states above the LUMO as there are no such bands.

Our work also shows the significance of considering the geometric structures of the doped QDs. Surface morphology is an integral part of QDs and plays a significant role in determining their LSPR properties and cannot be neglected by taking an ideal spherical geometry, as was done in ref. 15, for instance as some facets are polar and others are not. We found the polarization of the electric field can be related to the geometric structure to determine the direction with significant LSPR response, as shown in Fig. 8. For a wurtzite quantum dot of ZnO, the polarization along the **c** axis results in absorption spectral lines with narrower linewidth in contrast to the polarization along the **a** axis and a stronger electric dipole moment due to the polar faces along **c** axis.

Finally, we predict such a method can be used to study LSPRs of larger QDs of 4 nm or 5 nm, with longer bond lengths than that of ZnO, for instance CdSe. The peak of the spectral line for the uniform distribution in the 3.0 nm QD is around ~ .41 eV which is in the intermediate infrared range, which could be pushed to far infrared wavelength with larger QDs. Such studies can also be extended to study magneto-optics in diluted magnetic semiconductor QDs.

*Corresponding author: mogus.mochena@famu.edu

SUPPLEMENTARY MATERIAL

The supplementary material contains further simulation results and additional data thats support the conclusions reported here and can be obtained from corresponding author with reasonable request.

ACKNOWLEDGMENTS

D.D. and M.D.M were supported by the National Science Foundation Grant No. DMR-2013854. M.D.M. acknowledges computer time allocation (No. TG-DMR100055) for Stampede at Texas Advanced Computing Center from the Advanced Cyberinfrastructure Coordination Ecosystem: Services & Support (ACCESS) program, which is supported by National Science Foundation grants #2138259, #2138286, #2138307, #2137603, and #2138296. M.D.M. also acknowledges computer time at the carbon cluster of the Center for Nanoscale Materials at the Argonne National Laboratory and useful discussions with Dr. Pierre Darancet.

AUTHOR DECLARATIONS

Conflict of Interest
  The authors have no conflicts to disclose.

DATA AVAILABILITY

The data that support the findings of this study are available from the corresponding author upon reasonable request and are available within the article and its supplementary material.